\documentclass[aps,prl,twocolumn,superscriptaddress]{revtex4-2}
\usepackage{graphicx,xcolor}
\usepackage{amsmath,amssymb,mathtools,slashed}
\usepackage{tikz-feynman}[luatex=true]
\usepackage{feynmp-auto}
\usepackage{simpler-wick}
\usepackage{kantlipsum}
\usepackage{float}
\usepackage{multirow}
\usepackage{rotating}

\usepackage[T1]{fontenc} % if needed
\usepackage{etoolbox}
\usepackage{extarrows} 
\usepackage{graphicx}
\usepackage{dcolumn}
\usepackage{bm}
\usepackage{wasysym}
\usepackage{verbatim}

\usepackage{empheq}
\usepackage{multirow}
 
\usepackage{balance}

\usepackage{subfigure}
\usepackage[colorlinks=true,urlcolor=blue,linkcolor=blue]{hyperref}
\usepackage[capitalise]{cleveref}
\usepackage{siunitx}
\usepackage{enumerate}

\usepackage{array}
\usepackage{tcolorbox}
\usepackage{fontawesome}

\newcommand{\p}{\partial}

\newcommand{\bx}{{\bf{x}}}

\newcommand{\mH}{\mathcal{H}}

\newcommand{\bea}{\begin{eqnarray}}
\newcommand{\eea}{\end{eqnarray}}

\newcommand{\bq}{{\bf{q}}}

\usepackage{upgreek}

\newcommand{\mpl}{M_{\mbox{\tiny{Pl}}}}

\begin{document}

%=============================================================================
\title{%Gravitational Anomaly and Stochastic Chiral Fermion Production \\
Gravitational ABJ Anomaly, Stochastic Matter Production, and Leptogenesis}
%\title{Gravitational Chiral Anomaly and Stochastic Fermion Creation}
\date{\today}
\author{Azadeh Maleknejad}
\email{azadeh.maleknejad@kcl.ac.uk}
\affiliation{Department of Physics, King’s College London, Strand, London WC2R 2LS, UK}

%=============================================================================

\preprint{KCL-PH-TH/2024-74}

\begin{abstract}

Partially chiral stochastic gravitational wave backgrounds can arise from various processes in the early Universe. The gravitational ABJ anomaly links the chirality of the gravitational field to the chiral fermions, resulting in the stochastic generation of fermionic matter in the radiation era. We show that this mechanism can account for the entire dark matter relic density and discuss its potential as a leptogenesis scenario to explain the observed matter-antimatter asymmetry in the Universe.
Remarkably, this letter reveals that even if gravitational waves become unpolarized later, their temporary chirality during their generation process leaves a lasting imprint on the fermionic matter, preserved as a quantum remnant.

\end{abstract}

%=============================================================================
\maketitle
%=============================================================================

%=============================================================================
%\section{Introduction}
%=============================================================================
%
\begin{center}
    {\bf Introduction.}
\end{center}

\begin{figure}
    \centering
    \includegraphics[width=0.5\columnwidth]{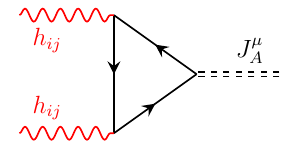}
    \caption{The gravitational ABJ anomaly:
$J^{\mu}_A$ is shown by a dashed black line, while gravitons are shown as wavy red lines.}
    \label{fig:diagram}
\end{figure}

Various processes in the early Universe can generate a partially chiral stochastic Gravitational Wave background (GWs), even if only transiently.  Some of the most 
prominent sources are: helical turbulence \cite{Kahniashvili:2020jgm}, magnetogenesis \cite{He:2021kah}, gauge fields in inflation \cite{Maleknejad:2012fw, Dimastrogiovanni:2012ew, Adshead:2013qp}, gauge prereheating \cite{Adshead:2015pva}, and modified gravity \cite{Alexander:2004us, Fu:2024ipa}. This induces a non-zero $R\tilde R$, a signature of chirality in the gravitational field that can generate fermions via the global gravitational anomaly in the Standard Model (SM) or beyond (BSM) \cite{Delbourgo:1972xb, Dowker:1977tp, Eguchi:1976db, AlvarezGaume:1983ig}. 
This phenomenon is the gravitational analog of the Adler-Bell-Jackiw (ABJ) anomaly  (see \cref{fig:diagram}). %\footnote{The global anomalies differ from gauge anomalies, which are forbidden by unitarity \cite{AlvarezGaume:1983ig}. } 
In the SM, the B-L current exhibits such an anomaly, enabling chiral GWs to generate a matter asymmetry when production occurs out of thermal equilibrium. This mechanism has been proposed in \cite{Alexander:2004us} as a leptogenesis scenario within the framework of modified gravity. It was later recognized that non-Abelian gauge fields in axion-inflation,  naturally produce chiral GWs \cite{Maleknejad:2012fw, Dimastrogiovanni:2012ew, Adshead:2013qp} and provide an intrinsic setting for gravitational leptogenesis in General Relativity \cite{ Maleknejad:2014wsa, Maleknejad:2016dci, Caldwell:2017chz, Alexander:2018fjp}. Notably, this setup predicts a chiral GWs background, detectable by future experiments  \cite{Komatsu:2022nvu}. Another quantum anomaly that can influence the dark matter relic abundance and baryon asymmetry during inflation is the chiral anomaly of a chiral SU(2) field \cite{Maleknejad:2020yys, Maleknejad:2020pec}. This mechanism can be more efficient than gravitational processes but relies on introducing new interactions with SM. In contrast, gravity is universal and unavoidable, making this effect an ever-present phenomenon.

 This raises two fundamental questions: I) Assuming a stochastic background of GWs generated during the radiation era, could this effect significantly contribute to the matter relic density and baryon asymmetry today, or is it more akin to Hawking radiation \cite{Hawking:1975vcx}, unavoidable yet negligible?
II) Does this process require the presence of chiral GWs today, or is the asymmetry solely during the transient phase sufficient? These questions are the focus of this work. We show that this mechanism can account for the present dark matter relic density and, in specific regions of parameter space, also explain the observed matter-antimatter asymmetry in the Universe. Furthermore, we find that even if GWs lose their polarization over time, their transient chirality during the generation process is sufficient for this phenomenon.

%=============================================================================
%\section{Fermions in an Expanding Universe}
%\label{Sec:fermion-FRW}
%=============================================================================
%

\begin{center}
    \textbf{Gravitational Anomaly: Breaking CP and Conformal Symmetry.}
\end{center}

The global gravitational anomaly of chiral fermions is 
\begin{align}
\label{G-anomaly-in}
\nabla_\mu J_A^\mu=\frac{N_{\chi}}{24(4\pi)^2}\langle R\tilde R \rangle,
\end{align}
where $J_A^\mu=\bar{\psi}\gamma^5\gamma^{\mu}\psi$ is the axial current, $N_{\chi}=N_L-N_R$ is the number of left- minus right-handed fermion species. For instance, in the SM with 3 left-handed neutrinos, $N_\chi=3$. Here $R\tilde R$ is the gravitational Pontryagin density defined as $   R\tilde R
\equiv\frac12\epsilon^{\lambda\mu\nu\xi}R_{\lambda\mu\rho\sigma}R_{\nu\xi}^{~~\rho\sigma}$ where $\epsilon^{\lambda\mu\nu\xi}$ is the totally antisymmetric tensor. That can be written as the divergence of the gravitational Chern-Simons 3-form as $\nabla_{\mu} K^{\mu} = R\tilde R$ where 
\begin{align}
K^{\mu}=\epsilon^{\mu\nu\lambda\sigma}\Gamma^{\beta}_{\nu\alpha}\big(\partial_{\lambda}\Gamma^\alpha_{\sigma\beta}+\frac23\Gamma^\gamma_{\lambda\beta}\Gamma^\alpha_{\sigma\gamma}\big).
\end{align}
Here $\{\Gamma^{\mu}_{\nu\lambda}\}$ is the Christoffel symbol. Since only parity-violating structures contribute to $R\tilde R$,  we exclude scalar perturbations and focus on GWs in the perturbed metric
\begin{align}
ds^2 = a^2(\tau) (-d\tau^2 + (\delta_{ij} + h_{ij}) dx^i dx^j),
\end{align}
where $\tau$ is the conformal time ($d\tau=a^{-1}dt$) and $h_{ij}$ is the GWs in the transverse traceless gauge. The unit normal vector of a cosmic observer experiencing $\tau$ as time is $n_{\mu}=(a(\tau),0,0,0)$. Consider a stochastic background of gravitational waves in the radiation era, expressed as %in terms of circular polarization states as
\bea
h_{ij}(\tau,\bx)=\sum_{s=\pm} \int d^3 q \, \hat{h}_{s,\bq}(\tau)\, e^{s}_{ij}(\textbf{q}) \, e^{i\bq.\bx},
\eea
where $e^{\pm}_{ij}(\bq)$ are the polarization tensors associated with the left- and right-handed polarization states of GWs. The operator  $\hat{h}_{s,\bq}(\tau)$ can be expanded as
\begin{align}
\hat{h}_{s,\bq}(\tau)= \hat{a}^{s}_{\bq}\, {\rm{h}}_{s,\bq}(\tau)+\hat{a}^{s \dag}_{-\bq} {\rm{h}}^*_{s, -\bq}(\tau),
\end{align}
where $\hat{a}^s_{\bq}$ are stochastic normalized variables satisfying  $\langle \hat{a}^s_{\bq} \hat{a}^{s'*}_{\bq'}\rangle = \delta_{ss'}\delta^{(3)}(\bq-\bq')$. 

The chiral gravitational charge can be defined as $\mathcal{K} = n_{\mu} K^{\mu}$, with its displacement given by
\begin{align}\label{eq:DeltaK-5---}
\Delta\mathcal{K}(\tau) \equiv \langle n_{\mu}K^{\mu}  (\tau) \rangle - \langle n_{\mu}K^{\mu}(\tau_{\text{in}})\rangle,
\end{align}
where $\tau_{\text{in}}$ represents the initial conformal time. That quantities the redistribution of chirality in the gravitational field, influenced by the coupling between spacetime curvature and the early cosmic Plasma. Up to quadratic order in cosmic perturbations, we obtain 
\begin{align}\label{eq:DeltaK-5-}
\Delta\mathcal{K} =\frac{4}{a^3}\sum_{s=\pm} s \!\int d^3q \, q
\big(\big|\p_{\tau} {\rm{h}}_{s,\bq}\big|^2-  q^2 \big|{\rm{h}}_{s,\bq}\big|^2 \big)\Big\vert_{\tau_{\text{in}}}^{\tau}.
\end{align}
This implies that a non-trivial  $\Delta\mathcal{K}$ requires: (i) a parity-violating evolution for GWs, and (ii) deviation from free-wave evolution during their propagation.

%=============================================================================
%\section{Fermion Production by Stochastic Gravitational Waves at 1-Loop}
%\label{Sec:fermion-1-loop}
%=============================================================================
% Assuming statistical isotropy of GWs,

Through the gravitational anomaly, cosmic perturbations induce spontaneous CP violation in fermions. Additionally, GWs break the conformal flatness of the FLRW geometry, inevitably violating the conformal symmetry of massless fermions in the cosmological background. This interplay gives rise to the anomalous production of fermions with an induced chiral charge defined as $\mathcal{Q}_A = n_{\mu} J^{\mu}_A$, with its displacement given by
\begin{align}\label{eq:Q5--}
 \Delta\mathcal{Q}_A(\tau) = \frac{N_{\chi}}{24(4\pi)} \, \Delta\mathcal{K}(\tau).
\end{align}
To evaluate this quantity, it is essential to model the stochastic gravitational wave (GW) background, which will be addressed in the next section.

%which computes the number density of chiral fermions created by parity-violating spacetime fluctuations.
%=============================================================================
%\section{Stochastic Gravitational Wave Backgrounds}
%=============================================================================
%
\begin{center}
    {\bf Stochastic Gravitational Wave Backgrounds.}
\end{center}

Here we present a phenomenological model for GWs background. We denote the conformal times marking the start and end of GWs production as $\tau_\text{in}$ and $\tau_*$, respectively.  This interval, referred to as the transient period, ends at $\tau_*$, which we define as the settling time. The GWs power spectrum at a given conformal time $\tau$ captures the combined effects of the GWs generation dynamics during the transient period, and the redshift induced by the Universe's expansion history. To disentangle these effects, we can decompose the GWs mode function as
\begin{align}
    \mathrm{h}_{s,\bq}(\tau) = a^{-1}(\tau) \, \mathcal{T}_s(\tau,q) \,  e^{-iq\tau} \, \mathrm{h}_{s,\bq,0},
    \label{eq:hsq-parameterization}
\end{align}
where $\mathrm{h}_{s\bq,0}$ is the GWs spectral amplitude of $s=\pm$ polarization state today at $\tau=\tau_0$, and $\mathcal{T}_s(\tau,q)$ is a transfer function that captures the dynamics of the GW production process and its formation during the transient phase.$[\tau_\text{in},\tau_*]$. We parameterize the transfer function as
\begin{align}
    \mathcal{T}_s(\tau,q) \approx \big( 1 - e^{-\pi \beta_s(\tau-\tau_\text{in})} \big),
    \label{eq:transfer-function}
\end{align}
where $\beta_s^{-1}$ is the characteristic time scale associated with the process that sources the $s$-polarization of GWs. For later convenience, we decompose $\beta_{\pm} $ as 
\begin{align}
    \beta_{\pm} = \beta (1 \pm b_{\chi}),
    \label{eq:Bpm}
\end{align}
where $b_{\chi}$ is a bounded real parameter ($\lvert b_{\chi}\rvert \leq 1$) that quantifies the chirality in the time evolution of GWs, capturing the difference in how the left- and right-handed modes evolve during the transient period. We assume that $\beta^{-1}$ is shorter than a Hubble time, i.e., $\beta/\mathcal{H}_* > 1$. Here $\mathcal{H}_* \equiv a(\tau_*) \, H(\tau_*)$, and $H(\tau)$ is the value of the Hubble parameter at a given conformal time. 

In the GWs literature, primordial GWs are commonly expressed in terms of their fractional cosmological energy density as observed today. %\footnote{The energy density of GWs is defined as $\rho_\text{gw}  = \frac{1}{32\pi G} \big\langle \dot{h}_{ij}\, \dot{h}^{ij}\big\rangle$ and its spectral energy density is $\Omega_\text{gw} = \frac{1}{\mpl^2} \frac{d\rho_\text{gw}}{3H^2 d\ln q} $.}
Furthermore, a commonly adopted assumption is the stochastic isotropy of the GW background, leading to the widely used expression
$\Omega_{\text{gw},0}(q) = \sum_{s=\pm}\frac{2\pi q^5}{3 H_0^2} \lvert \mathrm{h}_{s,\bq}(\tau_0) \rvert^2$. Here $H_0 \approx 10^{-60} \mpl$ is the present value of the Hubble constant. The energy density of each polarization of GWs is given as
\begin{align}
    \Omega_{\text{gw},0}^{\pm}(q) = \frac{1 \pm \chi(q)}{2}  \Omega_{\text{gw},0}(q),
    \label{eq:chi}
\end{align}
where $\chi(q)$ represents the degree of circular polarization (or Parity violation) of the GWs today, with $\lvert \chi(q) \rvert<1$.
%Adopting this convention, we obtain {\tcr{$\tau \gg \tau_*$}}
It is important to note that the degree of circular polarization in GWs is parametrized by two quantities: $b_{\chi}$, which is relevant only during the transient dynamics, and $\chi(q)$, which directly relates to the $V$ Stokes parameter observed in GWs today.
%=============================================================================
%\section{Dark Matter Relic Density}
%=============================================================================
%

\begin{center}
 {\bf Anomalous Fermion Production by Stochastic Chiral Gravitational Waves.}   
\end{center}

Now we are ready to compute the fermion creation by the gravitational anomaly. For $\tau\gg \tau_*$, the integrand in \cref{eq:DeltaK-5-} rapidly settles to a constant value, yielding
\begin{align}\label{eq:Q5-hh}
\big(\big|\p_{\tau} {\rm{h}}_{s,q}\big|^2-  q^2 \big|{\rm{h}}_{s,q}\big|^2 \big)\Big\vert_{\tau_{\text{in}}}^{\tau}  =  - \Big(\frac{\pi\beta_s}{a_{\text{in}}}\Big)^2  \lvert \mathrm{h}_{s,\bq}(\tau_0) \rvert^2.
\end{align}
 Consequently, fermion production predominantly occurs within the interval $[ \tau_{\text{in}},\tau_* ]$, effectively stopping afterward. Using \cref{eq:Q5-hh}, $\Delta\mathcal{Q}_A$ can be expressed in terms of the present-day GW power spectrum as
\begin{align}\label{eq:Q5}
\Delta\mathcal{Q}_A(\tau) & = \frac{N_{\chi}}{a^3(\tau)}   \!\int d\ln q \,\, n_A(q),
\end{align}
where the fermion spectral number density is
\begin{align}
n_A(q) & = -\frac{ \Omega_{\text{gw},0}(q)}{16 q} \big(\frac{ H_0\beta}{a_{\text{in}}}\big)^{\!2}   \big[  2b_{\chi} + (1+b_{\chi}^2)\chi(q)\!\big].
\label{eq:memory}
\end{align}
This result is particularly intriguing, as it suggests that even if the GW background since the settling time remains unpolarized (even if $\chi(q)=\Omega_{\text{gw}}^+(q,\tau) -\Omega_{\text{gw}}^-(q,\tau) =0 $ for $\tau> \tau_*$), differences in the time evolution of the circular modes during the transient dynamics ($b_{\chi} =\beta_+ -\beta_- \neq 0$) is enough to induce this phenomenon.

To compute the momentum integral in \cref{eq:Q5}, we require the present-day energy density of GWs, which varies depending on the specific early Universe scenario and requires complex numerical simulations to compute accurately. However, these results are typically well-approximated by simple analytical fitting functions. In this work, we adopt a broken power-law model for the net energy density of GWs today
\begin{align}
    \Omega_{\text{gw},0}(q) \approx
        \begin{cases}
            \Omega_\mathrm{p} \, \big( \frac{q}{q_\mathrm{p}} \big)^{m}
                &  q_\text{min} < q < q_\mathrm{p} , \\
            \Omega_\mathrm{p} \, \big( \frac{q}{q_\mathrm{p}} \big)^{-n}
                &  q_\mathrm{p} < q < q_\text{max} ,
                \end{cases}
    \label{eq:Omega-PT}
\end{align}
with  $m, n \geq 0$. This model describes a spectrum characterized by a peak, with the maximum spectral energy density, $\Omega_\mathrm{p}$, occurring at a characteristic momentum $q_\mathrm{p}$. The spectrum is governed by three key physical scales: at low frequencies, the horizon scale, $q_\text{min} = \mathcal{H}_*$ establishes a natural cutoff; at high frequencies, the spectrum extends to the smallest scales associated with the source, typically around $q_\text{max}\approx a_* T_*$ where $T_*$ represents the temperature of the hot early Universe plasma at $\tau_*$, and  $q_\mathrm{p}$ itself defines the characteristic scale of the source. \Cref{eq:Omega-PT} effectively models the results of simulations across various scenarios, such as phase transitions \cite{Caprini:2009yp, Durrer:2010xc} and primordial magnetic fields \cite{Caprini:2018mtu, RoperPol:2022iel}. Gravitational waves (GWs) produced during phase transitions exhibit distinct spectral features: at frequencies below the peak, the spectral index is typically $m \approx 3$, while at higher frequencies it ranges from $n \sim 1-4$, depending on the dominant GW production mechanism and the characteristics of the phase transition.

\begin{figure*}
    \begin{center}
    \includegraphics[height=4cm]{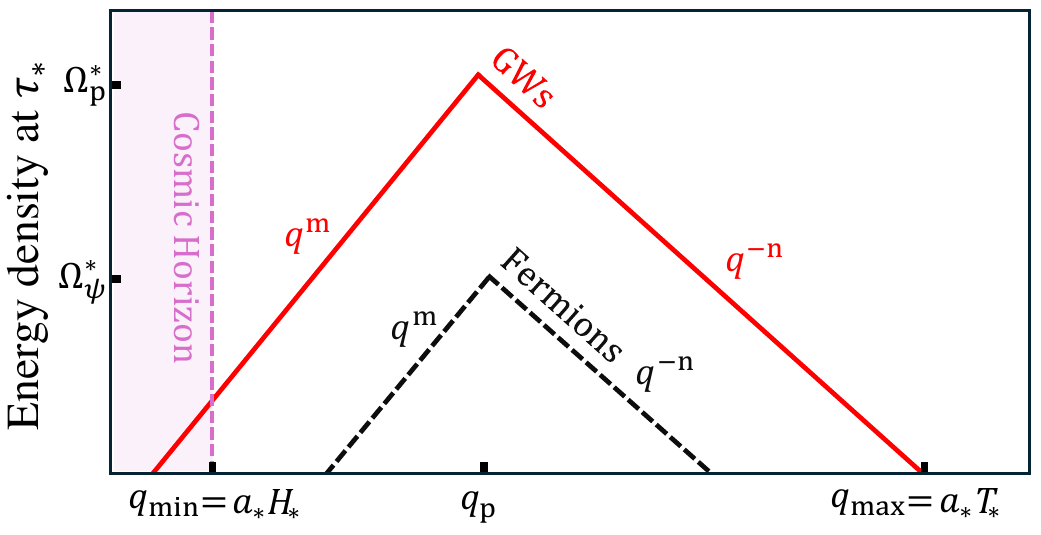} % Left plot
    \hfill
    \includegraphics[height=4cm]{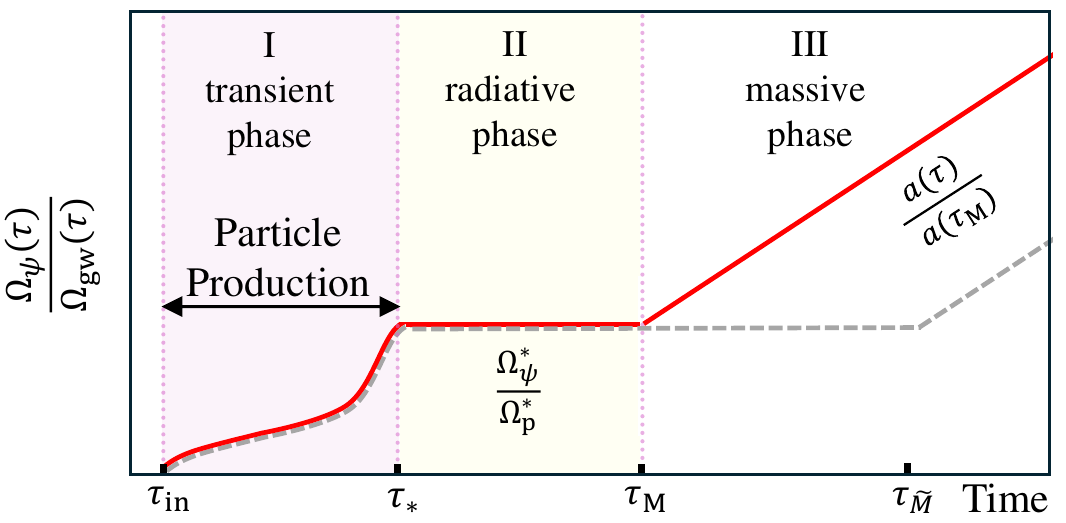} % Right plot
    \caption{ \textit{Left Panel:}  The spectral energy densities of GWs, $\Omega_{\text{gw},*}(q)$ (solid red), and fermions,  $\Omega_{\psi,*}(q)$ (dashed black), as functions of $q$, assuming $\chi(q)\approx \chi_0$. \textit{Right Panel:} The evolution of  $\frac{\Omega_{\psi}(\tau)}{\Omega_{\text{gw}}(\tau)}$. The solid (red) line shows the ratio 
over conformal time. The dashed line (gray) provides a reference, showing the evolution for a lighter fermion with mass $\tilde{M}<M$. }
    \label{fig:two_plots}
    \end{center}
\end{figure*}

For \cref{eq:Omega-PT}, the fermion spectral number density is 
\begin{align}
    n_{A}(q) \approx
        \begin{cases}
            \mathcal{N}_\mathrm{p}(q) \, q^3_\text{p} \, \big( \frac{q}{q_\mathrm{p}} \big)^{m-1}
                &  q_\text{min} < q < q_\mathrm{p} , \\
            \mathcal{N}_\mathrm{p}(q) \, q_\text{p}^3 \, \big( \frac{q}{q_\mathrm{p}} \big)^{-n-1}
                &  q_\mathrm{p} < q < q_\text{max} ,
                \end{cases}
    \label{eq:Omega-n}
\end{align}
where  $\mathcal{N}_\mathrm{p}(q)$ is a dimensionless quantity defined as
\begin{align}\label{eq:n-5}
\mathcal{N}_\mathrm{p}(q) = -\frac{ \Omega_{\text{p}}}{16} \big(\frac{ H_0\, \beta}{a_{\text{in}}\, q^2_\text{p}}\big)^{\!2}   \big[  2b_{\chi} + (1+b_{\chi}^2)\chi(q)\!\big].
\end{align}
 As a result, the spectral energy density of fermions is
\begin{align}
    \rho_{\psi}(q,\tau) = \frac{q\, \lvert N_{\chi} n_A(q)\rvert }{a^4(\tau)}.
    \label{eq:fermion-spectral-}
\end{align}
To ensure clarity and simplicity, we assume that $\chi(q)$ is nearly constant with respect to momentum
\begin{align}
\chi(q) \approx \chi_0.
\end{align}
Notably, in this case, the fermion and GW spectral energy densities share the same functional form (see left panel of \cref{fig:two_plots}).  For later convenience, we define 
\begin{align}
  \mathcal{B}_{\chi}  \equiv \frac14 \big[ 2b_{\chi} + (1+b^2_{\chi})\chi_0\big],
\end{align}
which is a dimensionless parameter that quantifies the chirality of the gravitational field, with $ \lvert\mathcal{B}_{\chi}\rvert <1$. Additionally, we take $m\approx 3$ and adopt the natural hierarchy of scales, $\mathcal{H}_* \ll q_\mathrm{p}\ll q_\mathrm{max} $.  Building on these assumptions, we find 
\begin{align}\label{eq:Q555}
\Delta\mathcal{Q}_A(\tau)  = - \frac{N_{\chi}}{8} \Big(\frac{q_\mathrm{p}  }{a(\tau)}\Big)^3 \big(\frac{ H_0}{a_{\text{in}}q_\mathrm{p} }\big)^{\!2}\Big(\frac{\beta}{q_\mathrm{p}}\Big)^2 \mathcal{B}_{\chi}\Omega_\mathrm{p},
\end{align}
which relates the number density of induced chiral fermions to the chirality of the gravitational field $\mathcal{B}_\chi$.

\begin{center}
{\bf Dynamics of Fermion Energy Density.}
\end{center}

In this section, we explore the evolution of fermion energy density $\rho_{\psi}(\tau)$, spanning the transient period to the present day. Here, we assume that the fermions, while effectively massless at the time of production $[\tau_\text{in},\tau_*]$, have non-negligible mass $M$ today. This mass may arise in two scenarios: the fermions possess a small initial mass $M \ll T_*$ from the start, or they remain massless through the phase transition and gain mass later via a Higgs mechanism at lower temperatures.

The evolution of the energy density exhibit three distinct phases: I) transient phase $[\tau_\text{in},\tau_*]$, characterized by GWs and fermion production, II) radiative phase $[\tau_*,\tau_M]$, during which both fermions and GWs remain massless and radiative, resulting in $\frac{\rho_{\psi}(\tau)}{\rho_\text{gw}(\tau)}= \text{const}$, and III) massive phase $(\tau>\tau_M)$, where fermions acquire mass, leading to $\frac{\rho_{\psi}(\tau)}{\rho_\text{gw}(\tau)}\propto a(\tau)$ (See right panel of \cref{fig:two_plots}).

At the end of the transient phase $(\tau=\tau_*)$, the peak of the GW spectral energy was significantly higher than its present-day value, given by $\Omega^*_\text{p}=a_*^{-4}\Omega_\textit{p} $. The peak of the fermion energy density at $\tau=\tau_*$ is
\begin{align}
    \Omega^*_{\psi} & = \frac{\lvert N_{\chi} \mathcal{B}_{\chi}\rvert  }{12 }  \big(\frac{ \beta} { \mH_*}\big)^{\!2} \big(\frac{a_*}{a_{\text{in}}}\big)^2 \, \big(\frac{H_*} {\mpl}\big)^{\!2} \, \Omega^*_\text{p},
\end{align}
where  $\mpl= 2.4 \times 10^{18}$ GeV is the reduced Planck mass. Furthermore, the ratio of fermion-GWs energy density remains constant throughout the radiative phase, where both behave as relativistic components. Given that $\mathcal{B}_\chi$ and $N_{\chi}$ are both order one quantities and $\frac{H_*}{\mpl} \lesssim 10^{-6}$, the backreaction of fermions on the gravitational background is negligible unless  $ \frac{ \beta} { \mH_*} \frac{a_*}{a_{\text{in}}} \frac{H_*/\mpl}{10^{-6}} \gtrsim 10^{6}$. Eventually, at $\tau=\tau_M$, the fermions become massive and their energy density at $\tau>\tau_M$ is
\begin{align}
\rho_{\psi}(\tau) = \frac{ 1}{a^3(\tau)} \frac{1}{a_M}  \frac{\lvert N_{\chi} \mathcal{B}_{\chi}\rvert}{12}  \big(\frac{ \beta} { \mH_*} \frac{a_*}{a_{\text{in}}} \frac{H_*} {\mpl}\big)^{\!2} \, \Omega_\text{p}\, \rho_c,
\end{align}
where $a_M=a(\tau_M)$, $\rho_c = 3H_0^2\mpl^2$, and we set the scale factor today as $a_0=1$. At this stage, fermions are non-relativistic and dilute like $a^{-3}$ while GWs remain relativistic. As a result, the initially negligible fermion to GW energy density increases with time and eventually can be very large.

\begin{center}
{\bf Dark Matter Relic Density.} 
\end{center}

\begin{figure}
    \centering
    \includegraphics[width=0.7\columnwidth]{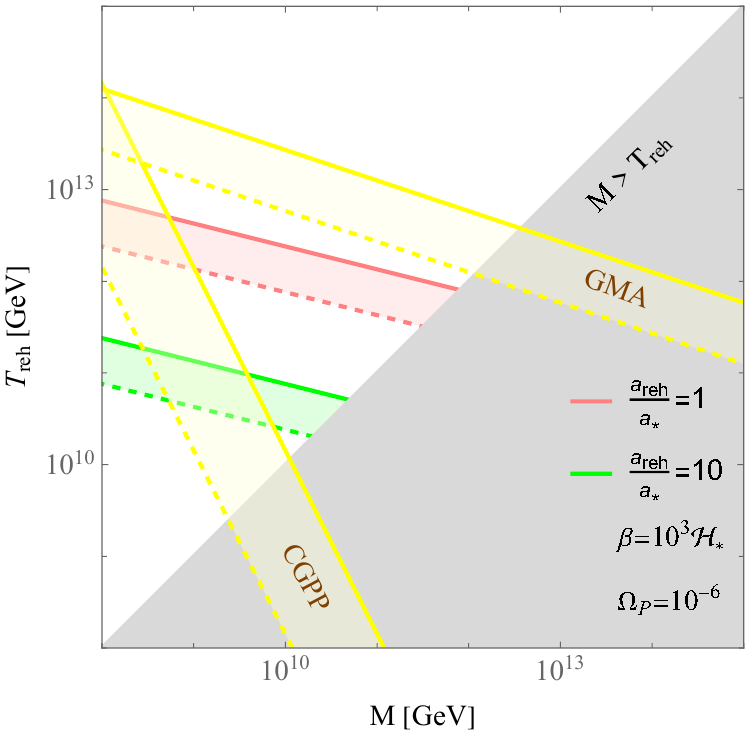}
    \caption{Gravitational ABJ-induced dark matter for a GW background with a broken power-law spectrum and $N_\chi=3$. Solid and dashed lines represent 100\% and 1\% fraction of the DM relic density today, respectively. The pink shaded region assumes $a_\text{ref}/a_*=1$, while the green shaded region represents $a_\text{ref}/a_*=10$. For comparison, the yellow regions highlight the parameter space where supermassive fermions achieve the relic density via conventional cosmological production (CGPP) \cite{Kolb:2017jvz, Ema:2019yrd, Kolb:2023ydq}, and graviton-mediated annihilation (GMA) \cite{Bernal:2018qlk, Clery:2021bwz}. The gray region represents fermions that are already massive at $T_\text{reh}$, making the gravitational anomaly mechanism inapplicable under the assumption $T_*\sim T_\text{reh}$.}
    \label{fig:DM}
\end{figure}

Here we assume the generated dark fermions are stable, making them viable cold dark matter candidates. %, and compute their contribution to the present-day dark matter relic density. 
Their fractional cosmological energy density today is
\begin{align}
\Omega_{\psi,0} =  \frac{M}{T_0} (\frac{g_{\text{eff},M}}{g_{\text{eff},0}})^{\frac13}  \frac{\lvert N_{\chi} \mathcal{B}_{\chi}\rvert}{12}  \big(\frac{ \beta} { \mH_*} \frac{a_*}{a_{\text{in}}} \frac{H_*} {\mpl}\big)^{\!2} \, \Omega_\text{p},
\end{align}
where  $T_0 = 2.3 \times 10^{-4}$ eV is the temperature of the universe today, $g_{\text{eff},0} = 3.38$, and $g_{\text{eff},M}= 106.75$,  the SM value above the electroweak scale. We find $\Omega_{\psi,0}$ as
\begin{align}
\Omega_{\psi,0} & \approx 0.3 \lvert N_{\chi} \mathcal{B}_{\chi}\rvert  \big(\frac{ \beta/\mH_*} { 10^3}\big)^{\!2} \big(\frac{a_*}{a_{\text{in}}}\big)^{\!2} \big(\frac{a_{\text{reh}}}{a_*}\big)^4 \big(\frac{M}{T_\text{reh}}\big)  \nonumber\\
& \times \big( \frac{T_\text{reh}} {10^{12} \text{GeV}}\big)^5\, \frac{\Omega_\text{p}}{10^{-6}},
\label{eq:DM}
\end{align}
and our results, shown in \cref{fig:DM}, demonstrate that this mechanism can account for the observed DM density across a broad range of high-scale DM masses and reheating temperatures. Interestingly, the efficiency of the gravitational ABJ-induced mechanism (CP-violating) is comparable to that of the coherent GWs scenario in the GW-induced freeze-in mechanism (CP-symmetric) in \cite{Maleknejad:2024ybn, Maleknejad:2024hoz}.

{\bf Stochastic Gravitational Leptogenesis.}

If SM leptons are out of thermal equilibrium during the transient phase, the Sakharov conditions are naturally satisfied in this setup, allowing this mechanism to generate a net lepton number. This requires reheating to occur after $\tau_*$, i.e. $a_*<a_\text{reh}$. In the SM with three left-handed neutrinos ($N_\chi=3$), the resulting lepton number density at $\tau=\tau_*$ can be read from \cref{eq:Q555}, $n_{\text{L,reh}} = \Delta \mathcal{Q}_A(\tau_\text{reh})$. Since it does not contribute to the baryonic sector, the gravitational anomaly serves as a $B-L$ production mechanism. Next, relying on the electroweak sphaleron processes, the generated primordial matter-antimatter
asymmetry in the lepton sector can transform into the baryonic sector. The baryon to photon number density today is related to its value at reheating as
\begin{align}
  \eta_{_{\text{B},0}} \equiv  \frac{n_\text{B,0}}{n_{\gamma,0}} = c_\text{sph}  \frac{g_{\text{eff},0}}{g_{\text{eff},\text{reh}}} \, \Big(\frac{n_\text{L,\text{reh}}}{n_{\gamma,\text{reh}}}\Big),
\end{align}
where $c_\text{sph}=\frac{28}{79}$ is sphaleron conversion factor.
 The photon number density at the time of reheating is $    n_{\gamma} = \frac{2\zeta(3)}{\pi^2} T_\text{reh}^3$, where $T_\text{reh}$ is the reheating temperature and $\zeta(3)=1.2$. 
The final baryon-to-photon ratio today is
\begin{align}
  \eta_{_{\text{B},0}} \approx   - 9\mathcal{B}_{\chi} \big(\frac{T_\text{reh}}{\mpl}\big)^3 \big(  \frac{a_\text{reh}}{a_*}\big)^3 \big(\frac{a_*}{a_{\text{in}}}\big)^2 \frac{\beta}{q_\mathrm{p} } \big(\frac{\beta/\mH_*}{10^3 } \big) \frac{\Omega_\mathrm{p}}{10^{-6}}. \label{eq:eta}
\end{align}
  An excess of matter over antimatter requires $\mathcal{B}_{\chi}<0$. Considering $\mathcal{B}_{\chi}\sim -1$ and $\beta \sim q_\text{p}$, this mechanism can account for the observed baryon-antibaryon asymmetry today, $\eta_{_{\text{B},0}}\approx 6 \times 10^{-10}$ \cite{Planck:2015fie},  provided 
  \begin{align}
      \big(\frac{T_\text{reh}}{\mpl}\big)^3 \big(  \frac{a_\text{reh}}{a_*}\big)^3 \big(\frac{a_*}{a_{\text{in}}}\big)^2  \big(\frac{\beta}{10^3\mH_* } \big) \frac{\Omega_\mathrm{p}}{10^{-6}} \sim 10^{-10}.
  \end{align}
The values of $\frac{a_\text{reh}}{a_*}$ and $\frac{a_*}{a_{\text{in}}}$ depend on the specifics of the phase transition and the reheating efficiency and can vary. \footnote{For a phenomenological model of reheating, see App. S3 in \cite{Maleknejad:2020yys}.} For simplicity and clarity, let us adopt typical values $\frac{a_\text{reh}}{a_*}\sim\frac{a_*}{a_{\text{in}}}\sim 3$ and $\beta \sim 10^3 \mH_*$. Interestingly, successful baryogenesis via the gravitational ABJ anomaly predicts a gravitational wave background near the BBN upper limit, $\Omega_\text{p}\sim 10^{-6}$,  alongside a reheating temperature approaching the Cosmic Microwave Background (CMB) upper bounds, $T_\text{reh}\sim 10^{14} \text{GeV}$.

%=============================================================================
%\section{Outlook}
%\label{Sec:conclusion}
%=============================================================================

\begin{center}
{\bf Summary and Outlook.}
\end{center}

In this letter, we have studied the gravitational ABJ anomaly (\cref{fig:diagram}) in the radiation era of the early Universe and its profound role in generating SM and dark sector fermions. We adopted phenomenological models for the evolution of GWs, including a broken power-law spectrum for their energy density (\cref{eq:Omega-PT}) and two chirality (parity violation) parameters:  the transient chirality  GWs during their generation process (\cref{eq:Bpm}), and the degree of circular polarization of GWs today (\cref{eq:chi}).

Notably, we found that even if GWs become unpolarized later, their transient chirality can produce fermions via the gravitational ABJ anomaly (\cref{eq:memory}). This cosmological phenomenon parallels similar effects in flat space, such as the chiral magnetic effect \cite{Kharzeev:2013ffa} and the chiral memory effect \cite{Maleknejad:2023nyh}. 
The spectral energy densities of the induced fermions and GWs share the same functional form (\cref{eq:fermion-spectral-} and the left panel of \cref{fig:two_plots}). The evolution of fermions has three distinct phases: (I) a transient phase, (II) a radiative phase, and (III) the massive phase (the right panel of \cref{fig:two_plots}). We showed that this mechanism can account for the present dark matter relic density (\cref{eq:DM}) and, in specific regions of parameter space, explain the observed matter-antimatter asymmetry in the Universe (\cref{eq:eta}). % Intriguingly, successful baryogenesis via the gravitational ABJ anomaly predicts a GWs background near the BBN upper limit, $\Omega_\text{p}\sim 10^{-6}$, and a reheating temperature approaching the Cosmic Microwave Background (CMB) upper bounds, $T_\text{reh}\sim 10^{14} \text{GeV}$. 

%=============================================================================

\begin{acknowledgments}
{\bf{Acknowledgments:} } %I am grateful to Dmitri Kharzeev, Eiichiro Komatsu, and Joachim Kopp for helpful discussions.

I am grateful to Dmitri Kharzeev, Eiichiro Komatsu, and Joachim Kopp for their valuable collaboration on related topics, which inspired this research.
My work is supported by the Royal Society University Research Fellowship, Grant No. RE22432. 
\end{acknowledgments}

%=============================================================================

%\include{G-Stochastic/triangle-diagram}

\bibliographystyle{apsrev4-1}
\bibliography{ref.bib}

%=============================================================================

\end{document}